\documentclass[12pt,a4paper]{amsart}
\usepackage[utf8]{inputenc}
\usepackage[T1]{fontenc}
\usepackage[english]{babel} 
\usepackage{amsmath, amssymb, amsthm}
\usepackage{geometry}
\usepackage{graphicx}
\usepackage{hyperref}
\usepackage{xcolor}
\usepackage{enumitem}
\usepackage{mathtools}
\usepackage{csquotes}
\usepackage{biblatex} 
\usepackage{babel}

\newcommand{\F}{\mathbb{F}}

\newcommand{\cO}{\mathcal{O}}

\newcommand{\cH}{\mathcal{H}}
\newcommand{\CL}{\mathcal{C}_\mathcal{L}}
\newcommand{\cE}{\mathcal{E}}

\def\cH{\mathcal{H}}
\def\cL{\mathcal{L}}

\def\Z{\mathbb{Z}}

\DeclareMathOperator{\lmd}{lmd}

\hypersetup{
    colorlinks=true,
    linkcolor=blue,
    filecolor=magenta,      
    urlcolor=blue,
    citecolor=blue,
}

\theoremstyle{plain}
\newtheorem{theorem}{Theorem}[section]
\newtheorem{lemma}[theorem]{Lemma}
\newtheorem{proposition}[theorem]{Proposition}
\newtheorem{corollary}[theorem]{Corollary}

\theoremstyle{definition}

\newtheorem{example}[theorem]{Example}

\theoremstyle{remark}
\newtheorem{remark}[theorem]{Remark}

\begin{document}

\title[LCP of MDS Elliptic Codes]{Linear Complementary Pairs of MDS Codes from Elliptic Curves}

\author{Adler Marques}
\address{Instituto de Matemática, Universidade Federal do Rio de Janeiro, Rio de Janeiro, Brasil}
\email{adler@im.ufrj.br}

\date{\today}

\begin{abstract}
In this paper, we investigate the Linear Complementary Pairs (LCPs) of Maximum Distance Separable (MDS) codes that arise from elliptic curves. We establish geometric conditions for LCPs of elliptic codes and prove that they cannot consist entirely of MDS codes when evaluated over all rational points of the curve. To address this, we construct LCPs of MDS elliptic codes by restricting evaluation divisors to subgroup cosets. We apply this method to obtain explicit infinite families of LCPs of MDS codes over Mordell and supersingular elliptic curves, including large-length constructions over quadratic extensions.
\end{abstract}

\maketitle

\section{Introduction}
\label{sec:intro}
A linear complementary pair (LCP) of codes over a finite field \(\mathbb{F}_q\) is a pair \((\mathcal{C},\mathcal{D})\) of linear codes of length \(n\) such that \(\mathcal{C}\oplus\mathcal{D}=\mathbb{F}_q^n\). In the special case \(\mathcal{D}=\mathcal{C}^\perp\), the code \(\mathcal{C}\) is called a linear complementary dual (LCD) code, a notion introduced by Massey \cite{massey1992}. Such pairs form the algebraic foundation of masking schemes against side-channel attacks (SCA) and fault injection attacks (FIA) \cite{carlet2014, carlet-guilley2015, bhasinetal2015}, with the security of the resulting implementation determined by the security parameter $\min\{ d(\mathcal{C}),\,d(\mathcal{D}^\perp)\}$. 

For the security parameter to be maximally effective, both \(\mathcal{C}\) and \(\mathcal{D}^\perp\) must have large minimum distances. The ideal scenario occurs when both are maximum distance separable (MDS) codes, attaining the Singleton bound with equality. Since MDS codes are optimal for their length and dimension, an LCP of MDS codes provides the strongest possible security parameter. However, constructing such pairs remains mathematically challenging.

Algebraic geometry (AG) codes offer a natural framework for constructing LCPs of codes \cite{BDM2024, CMQ2025, CLM2024}. Elliptic curves (genus \(g=1\)) are particularly appealing because their group structure and Riemann-Roch spaces provide precise control over code parameters. The study of MDS elliptic codes has a rich history: Munuera \cite{munuera93} gave the first bound on their length, Li, Wan, and Zhang \cite{LiWanZhang2015} conjectured that the maximal length of \(q\)-ary MDS elliptic codes is close to \(q/2\), and Han and Ren \cite{HanRen2023} confirmed this for a wide range of dimensions. Recently, Chen, Hu, Huang, and Zhao \cite{chen2026} resolved the remaining open cases, determining the exact maximal length \(\mathrm{MEC}(k,q)\) for all admissible \(k\). Constructively, Wang, Liu, Luo, and Zhai \cite{Wang2026} built new families of non-Reed--Solomon MDS codes from elliptic curves, while iso-dual MDS elliptic codes and their connections to quantum error-correcting codes were explored by Zhu and Zhao \cite{Zhu2026}. Nevertheless, evaluating a code over an entire subgroup \(\cH\) of an elliptic curve introduces the zero-sum relation $\sum_{P\in \cH}P=\cO$, which forces linear dependencies among the coordinates and destroys the MDS property. This rigidity explains why existing LCP constructions over prime fields are typically limited to shorter lengths or to almost-MDS codes.

We address this obstruction by constructing an evaluation divisor over a union of subgroup cosets from which a single low-order rational point \(P_0\) is excluded. This strategic approach breaks the zero-sum relation, enabling the explicit construction of LCPs of elliptic codes where both constituent codes are strictly MDS.

We begin in Section \ref{sec:preliminaries} by reviewing the necessary background on function fields, Riemann--Roch spaces, and elliptic curves. In Section \ref{sec:LCP}, we use the Abel--Jacobi theorem to establish sufficient geometric conditions for LCPs of elliptic codes and prove a fundamental MDS limitation regarding evaluations over all rational points. This motivates our multi-coset construction and the explicit congruence conditions required to guarantee the MDS property. Section \ref{sec:explicit_examples} illustrates these theoretical results by generating explicit infinite families of LCPs of MDS codes. We apply the construction to prime fields, specifically using Mordell curves for \(p \equiv 5 \text{ or } 11 \pmod{18}\) and the supersingular curve \(y^2=x^3+x\) for \(p \equiv 3 \pmod 8\), and extend the method to supersingular curves over $\mathbb{F}_{p^2}$ with group structure $\mathbb{Z}_{p+1} \oplus \mathbb{Z}_{p+1}$ to achieve large-length LCPs of  MDS codes of length $n=s(p+1)-1$. Finally, Section \ref{sec:conclusion} concludes the paper and outlines directions for future research.

\section{Preliminaries}
\label{sec:preliminaries}
In this section, we present the fundamental concepts that will serve as the basis for the remainder of this paper. We briefly review the theory of linear codes and linear complementary pairs of codes, followed by the basics of elliptic curves and finally introduce algebraic geometry elliptic codes. For more general results and further details on these topics, we refer the reader to \cite{guruswami2025essential, silverman2009, stichtenoth2009algebraic}.

\subsection{Linear Codes and Linear Complementary Pairs}

A linear code $\mathcal{C}$ of length $n$ over a finite field $\F_q$ is a vector subspace of $\F_q^n$. The dimension of $\mathcal{C}$ as an $\F_q$-vector space is denoted by $k$. The Hamming weight of a vector $v \in \F_q^n$ is the number of its non-zero coordinates, and the minimum distance $d$ of $\mathcal{C}$ is the minimum Hamming weight of its non-zero vectors. A linear code with length $n$, dimension $k$, and minimum distance $d$ is referred to as an $[n,k,d]_q$-code.

The parameters of a linear code satisfy the Singleton bound, which states that $d \le n - k + 1$. Codes achieving this bound, meaning $d = n - k + 1$, are called Maximum Distance Separable (MDS) codes. 

For two vectors $u = (u_1, \dots, u_n)$ and $v = (v_1, \dots, v_n)$ in $\F_q^n$, their standard Euclidean inner product is $\langle u, v \rangle = \sum_{i=1}^n u_i v_i$. The dual code of $\mathcal{C}$, denoted by $\mathcal{C}^\perp$, is the orthogonal complement of $\mathcal{C}$ with respect to this inner product:
\[
\mathcal{C}^\perp = \{ u \in \F_q^n \mid \langle u, v \rangle = 0 \text{ for all } v \in \mathcal{C} \}.
\]
If $\mathcal{C}$ is an $[n,k,d]_q$-code, its dual $\mathcal{C}^\perp$ is an $[n, n-k, d^\perp]_q$-code. It is a well-known property that the dual of an MDS code is also an MDS code.

Two linear codes $\mathcal{C}$ and $\mathcal{D}$ of length $n$ over $\F_q$ form a Linear Complementary Pair (LCP) if they satisfy the direct sum decomposition
\[
\mathcal{C} \oplus \mathcal{D} = \F_q^n.
\]
Equivalently, the pair $(\mathcal{C}, \mathcal{D})$ is an LCP of codes if and only if $\dim(\mathcal{C}) + \dim(\mathcal{D}) = n$ and their intersection is trivial, that is, $\mathcal{C} \cap \mathcal{D} = \{0\}$.

\subsection{Elliptic Curves} 

Let $\mathcal{E}/\mathbb{F}_q$ be an elliptic curve with identity element $\mathcal{O}$. We identify rational points of $\cE$ with rational places of the function field $\F_q(\cE)$. The set of rational points $\cE(\F_q)$ is a finite abelian group. We write $\oplus$ for the group law on $\cE(\F_q)$, $\ominus$ for the corresponding inverse operation and $+$ for the addition of divisors. For $P \in \cE(\F_q)$ and $m \in \mathbb{Z}$, the symbol $[m]P$ denotes scalar multiplication in $\cE(\F_q)$, while $mP$ denotes the divisor obtained by adding the place $P$ with multiplicity $m$.

By the Hasse Theorem, the number $|\cE(\F_q)|$ of $\F_q$-rational points of $\cE$ satisfies $|\cE(\F_q)| = q+1-t$, for some integer $t$ with $|t| \le 2\sqrt{q}$. It is also well-known that there are integers $n_1, n_2 \ge 1$ with $n_1 \mid n_2$ and $n_1 \mid (q-1)$ such that
\[
\cE(\F_q) \cong \Z/n_1\Z \oplus \Z/n_2\Z.
\]
For a description of all groups that can occur as $\cE(\F_q)$, see \cite{Voloch1988}.

For a rational divisor $D=\sum_P m_P P$ on $\cE$, its group sum is defined by
\[
    \sigma(D) = \bigoplus_P [m_P]P \in \cE(\F_q).
\]
The Abel-Jacobi Theorem states that the degree zero divisor class group is isomorphic to the group $\cE(\F_q)$ via the map $[D] \mapsto \sigma(D)$ \cite[Proposition 3.4]{silverman2009}. Under this isomorphism, a degree zero divisor is principal if and only if it maps to the identity element $\cO$. Since the difference $D_1 - D_2$ of two divisors of the same degree has degree zero, we obtain the equivalence
\[
D_1 \sim D_2 \iff \sigma(D_1) = \sigma(D_2).
\]

 For a divisor $A \in \operatorname{Div}(\cE)$, the Riemann-Roch space $\cL(A)$ associated with $A$ is the set 
\[
\cL(A)=\{z \in \F_q(\cE) \mid \operatorname{div}(z) \ge -A \} \cup \{0\}.
\]
From the Riemann-Roch Theorem, since $\cE$ has genus $g=1$, the dimension $\ell(A) := \dim_{\F_q} \cL(A)$ is given by
\[
\ell(A)= \deg(A)+\ell(K-A),
\]
where $K \in \operatorname{Div}(\cE)$ is a canonical divisor of $\cE$. In particular,
\begin{equation}
    \label{eq:dim_L(A)_g=1}
\ell(A)=\begin{cases}
    \deg(A), & \mbox{if } \deg(A) >0, \\
    1 , & \mbox{if } \deg(A)=0 \mbox{ and } A \mbox{ is a principal divisor,} \\
    0, &\mbox{otherwise.}
\end{cases}
\end{equation}
Recall that a divisor $A$ on $\cE$ is called non-special if $\ell(A)=\deg(A)$; equivalently, $\ell(K-A)=0$ for any canonical divisor $K \in \operatorname{Div}(\cE)$. Thus, degree-zero divisors are non-special if and only if they are non-principal.

\subsection{Elliptic Codes}

We now construct a specific class of linear codes using elliptic curves. Let $P_1, \dots, P_n \in \cE(\F_q)$ be $n$ distinct $\F_q$-rational points of $\cE$ and consider the divisor $D := P_1 + \cdots + P_n$. For a rational divisor $G$ on $\cE$ whose support is disjoint from that of $D$ and $0 < \deg(G) < n$, the elliptic algebraic geometric code (or simply, elliptic code) $\CL(D,G)$ associated with $G$ and $D$ is
\[
\CL(D,G) = \{ (f(P_1), \dots, f(P_n)) \mid f \in \cL(G) \} \subseteq \F_q^n.
\]
We have that $\CL(D,G)$ is an $[n,k,d]_q$-code, where
\begin{equation}
\label{eq:ParametersEllipCodes}        
k=\deg(G) \quad \mbox{and} \quad d=n-k \quad \mbox{or} \quad d=n-k+1.
\end{equation}
Consequently, by the definitions above, elliptic codes are either MDS $[n,k,n-k+1]_q$-codes or almost-MDS (or near-MDS) $[n, k, n-k]_q$-codes.

The following characterizes the MDS property for elliptic codes \cite{munuera93}.

\begin{lemma}
\label{lemma:MDS_characterization}
    Let $D=\sum_{i=1}^n P_i$ be the sum of $n$ distinct rational points of $\cE$, and let $G$ be a divisor with support disjoint from that of $D$. Suppose that $0 < \deg G < n$, and put $k := \deg G$. Then, $\CL(D, G)$ is an MDS code if and only if there are no $k$ distinct points $P_{i_1}, \dots, P_{i_k} \in \operatorname{Supp}(D)$ such that
    \[
    \sigma(G) = P_{i_1} \oplus \cdots \oplus P_{i_k}.
    \]
    Equivalently, $\CL(D, G)$ is almost-MDS if and only if there exist $k$ distinct points $P_{i_1},\ldots,P_{i_k} \in \operatorname{Supp}(D)$ such that \[ G\sim P_{i_1}+\cdots+P_{i_k}. \]
\end{lemma}

\subsection{LCP of AG Codes}
Constructing an LCP of codes requires two codes with trivial intersection and complementary dimensions. For AG codes, we translate these linear algebraic properties into geometric conditions on Riemann-Roch spaces by manipulating their underlying divisors.

Let $G$ and $H$ be two divisors on $\cE$, given by $G = \sum m_P P$ and $H = \sum n_P P$. The greatest common divisor ($\gcd$) and the least multiple divisor ($\operatorname{lmd}$) of $G$ and $H$ are defined naturally as
\[
\gcd(G, H) = \sum_P \min(m_P, n_P)P \quad \text{and} \quad \operatorname{lmd}(G, H) = \sum_P \max(m_P, n_P)P.
\]
From the definition of Riemann-Roch spaces, it holds $\cL(G) \cap \cL(H) = \cL(\gcd(G, H))$. By evaluating these spaces at $D$, we obtain the inclusion
\begin{equation}
\label{eq:inclusion_elliptic}
\CL(D, \gcd(G, H)) \subseteq \CL(D, G) \cap \CL(D, H).
\end{equation}
The reverse inclusion does not hold in general. A codeword in $\CL(D, G) \cap \CL(D, H)$ comes from functions $f \in \cL(G)$ and $h \in \cL(H)$ that evaluate to the same vector on $D$. This implies that their difference vanishes on $D$, meaning $f - h \in \cL(\operatorname{lmd}(G, H) - D)$. Thus, to determine the intersection of the codes, it is necessary to analyze the Riemann-Roch space of $\operatorname{lmd}(G, H) - D$. This algebraic behavior motivates the sufficient conditions for an LCP of AG codes, as established by Bhowmick, Dalai and Mesnager \cite{BDM2024}.

\begin{theorem}\cite[Theorem~3.5]{BDM2024} 
\label{thm:Thm3.5BDM24}
    Let $\mathcal{C}_\cL(D, G)$ and $\mathcal{C}_\cL(D, H)$ be two AG codes over $\mathbb{F}_q$ of length $n$ with genus $g > 0$, such that $2g-2 < \deg(G), \, \deg(H) < n$. Then $\left( \mathcal{C}_\cL(D, G), \, \mathcal{C}_\cL(D, H) \right)$ is an LCP of codes if:
    \begin{enumerate}
        \item $\ell(G) + \ell(H) = n$, and
        \item $\gcd(G,H)$ and $\operatorname{lmd}(G,H)-D$ are both non-special divisors of degree $g-1$.
    \end{enumerate}
\end{theorem}

\section{LCP of MDS Elliptic Codes}
\label{sec:LCP}

We now focus on the construction and analysis of linear complementary code pairs obtained specifically from elliptic curves. 

\subsection{General Conditions and MDS Limitations}

By applying Theorem \ref{thm:Thm3.5BDM24} to the case where the genus is $g=1$, the geometric conditions simplify substantially. For an elliptic curve, the degree bounds become $0 < \deg(G), \deg(H) < n$, which ensures that $\ell(G) = \deg(G)$ and $\ell(H) = \deg(H)$. Furthermore, as seen in Equation \eqref{eq:dim_L(A)_g=1}, a degree-zero divisor is non-special if and only if it is non-principal. Combining these facts with the Abel-Jacobi Theorem yields practical sufficient conditions for LCPs of elliptic codes.

\begin{corollary}
\label{cor:LCP_elliptic}
    Let $G$ and $H$ be divisors on an elliptic curve $\cE$ whose supports are disjoint from $D = P_1 + \cdots + P_n$. Suppose $0 < \deg(G), \deg(H) < n$. Then $\left( \CL(D, G), \, \CL(D, H) \right)$ is an LCP of elliptic codes if the following conditions hold:
    \begin{enumerate}[label=(\roman*)]
        \item $\deg(G) + \deg(H) = n$;
        \item $\deg(\gcd(G,H)) = 0$ and $\sigma(\gcd(G,H)) \neq \cO$; 
        \item $\deg(\operatorname{lmd}(G,H)) = n$ and $\sigma(\operatorname{lmd}(G,H)) \neq \sigma(D)$.
    \end{enumerate}
\end{corollary}

A fundamental question is whether an LCP of codes can be formed entirely by MDS codes. The following result establishes that if an elliptic curve does not admit MDS codes of a given low dimension, then any LCP of elliptic codes derived from it that contains a code of that dimension will be completely without MDS codes.

\begin{proposition}
    Let $\cE/\mathbb{F}_q$ be an elliptic curve and $D = P_1 + \cdots + P_n$ be the sum of $n$ distinct rational points of $\cE$. Let $\alpha$ be an integer such that $0 < \alpha < n$. Suppose that for every divisor $G' \in \operatorname{Div}(\cE)$ whose support is disjoint from $D$ and which satisfies $0 < \deg(G') < \alpha$, the elliptic code $\CL(D, G')$ is not MDS. If $(\CL(D,G), \CL(D,H))$ is an LCP of elliptic codes with $0 < \deg(G) < \alpha$, then neither $\CL(D,G)$ nor $\CL(D,H)$ is an MDS code.
\end{proposition}

\begin{proof}
    Let $(\CL(D, G), \CL(D, H))$ be an LCP of elliptic codes such that $0 < \deg G < \alpha$. By the initial assumption, since $0 < \deg G < \alpha$, we have $\dim \CL(D, G)= \deg G$ and $\CL(D, G)$ is necessarily not an MDS code. 
    
    Since $\CL(D,G) \oplus \CL(D,H) = \mathbb{F}_q^n$, we have
    \[
    \dim \CL(D, H) = n - \dim \CL(D,G) = n - \deg(G).
    \] 
     It follows that $\deg(H) = n - \deg(G)$.

    Suppose, by contradiction, that $\CL(D, H)$ is an MDS code. It is well-known that the dual of an MDS code is also MDS. Recall that the dual code $\CL(D, H)^\perp$ is the elliptic code $\CL(D, H^\perp)$, where $H^\perp := D - H + \operatorname{div}(\eta)$ for some suitable differential $\eta$ on $\cE$. Since the divisor of a differential on an elliptic curve has degree $2g - 2 = 0$, the degree of $H^\perp$ is given by
    \[
    \deg(H^\perp) = \deg(D) - \deg(H) + \deg(\operatorname{div}(\eta)) = n - (n - \deg(G)) = \deg(G).
    \]
   Consequently, we obtain $0 < \deg(H^\perp) < \alpha$. By hypothesis, this condition forces the elliptic code $\CL(D, H^\perp)$ to be non-MDS, contradicting our assumption. Hence, $\CL(D, H)$ cannot be MDS.
\end{proof}

\begin{remark}
\label{rem:MDS_free}
    The assumption that no code $\CL(D,G')$ of dimension $k < \alpha$ is MDS restricts the evaluation divisor $D$. By Lemma \ref{lemma:MDS_characterization}, this hypothesis requires that, for every $1 \le k < \alpha$, the subset sum map from the $k$-subsets of $\operatorname{Supp}(D)$ to $\cE(\F_q)$, given by $\{Q_1, \dots, Q_k\} \mapsto Q_1 \oplus \cdots \oplus Q_k$, must be surjective.  This implies $\binom{n}{k} \ge \vert\cE(\F_q)\vert$. If $\alpha \ge 2$, setting $k=1$ yields $n \ge \vert\cE(\F_q)\vert$. Since $\operatorname{Supp}(D) \subseteq \cE(\F_q)$ implies $n \le \vert\cE(\F_q)\vert$, we obtain $n = \vert\cE(\F_q)\vert$. Thus, the non-existence of MDS codes for dimensions $k < \alpha$ occurs only when evaluating over all $\F_q$-rational points of $\cE$.
\end{remark}

\subsection{Constructions of LCPs of MDS Elliptic Codes}

By Remark \ref{rem:MDS_free}, constructing an LCP of MDS elliptic codes requires the evaluation divisor $D$ to be supported on a proper subset of $\cE(\F_q)$. We provide explicit constructions of such pairs by forming $D$ from subgroup cosets, excluding at least two rational points of $\cE$.

Let $\mathcal{H} \leq \cE(\F_q)$ be a subgroup of order $m$, and let $P_0 \in \cE(\F_q)$ be such that $\langle P_0 \rangle \cap \mathcal{H} = \{ \cO \}$. Let $\Lambda = \{\lambda_1, \dots, \lambda_s\}$ be a set of $s$ distinct non-zero integers modulo $\operatorname{ord}(P_0)$ with $\lambda_1 = 1$. For each $1 \le j \le s$, define the coset $\mathcal{H}_j = [\lambda_j]P_0 \oplus \mathcal{H}$. The condition on $\Lambda$ ensures that $\cO \notin \mathcal{H}_j$ for all $j$. 

Let $U = \bigcup_{j=1}^s \mathcal{H}_j$ and define the evaluation divisor as
\[
D = \sum_{P \in U \setminus \{P_0\}} P.
\]
Let $n = \deg(D) = s m - 1$. The number of points each coset contributes to $\operatorname{Supp}(D)$ is given by $m_1 = m - 1$ and $m_j = m$ for $2 \le j \le s$. Finally, let $S_\Lambda = \sum_{j=1}^s \lambda_j$. By construction, $\operatorname{Supp}(D)$ is disjoint from $\{P_0, \cO\}$.
\begin{theorem}
\label{thm:LCP_elliptic_multicoset}
With the notation above, let $a$ and $b$ be positive integers satisfying $0 < b-a < n$, and assume that
\[
    a \not\equiv 0 \quad \mbox{and} \quad b \not\equiv 0 \pmod{\operatorname{ord}(P_0)}.
\]
Furthermore, assume the following conditions hold:
\begin{enumerate}
    \item $b \not\equiv m S_\Lambda - 1 \pmod{\operatorname{ord}(P_0)}$.
    \item For every sequence of integers $(k_1, \dots, k_s)$ such that $0 \le k_j \le m_j$ and $\sum_{j=1}^s k_j = b-a$, 
    \[ 
    b \not\equiv \sum_{j=1}^s k_j \lambda_j \pmod{\operatorname{ord}(P_0)}. 
    \]
    \item For every sequence of integers $(\ell_1, \dots, \ell_s)$ such that $0 \le \ell_j \le m_j$ and $\sum_{j=1}^s \ell_j = a+n-b$, 
    \[ 
    a \not\equiv \sum_{j=1}^s \ell_j \lambda_j \pmod{\operatorname{ord}(P_0)}. 
    \]
\end{enumerate}
Then, the pair of codes
\[
    \big(\CL(D, aP_0+(n-b)\cO), \ \CL(D,bP_0-a\cO)\big)
\]
forms an LCP of MDS elliptic codes with parameters $[n, a+n-b, b-a+1]_q$ and $[n, b-a, n-b+a+1]_q$, respectively.
\end{theorem}

\begin{proof}
Let $G = aP_0 + (n-b)\cO$ and $H = bP_0 - a\cO$. Note that $\deg(G) = a+n-b$ and $\deg(H) = b-a$. Since $0 < b-a < n$, we have $0 < \deg(G), \deg(H) < n$. Furthermore, $\operatorname{Supp}(G) \cup \operatorname{Supp}(H) \subseteq \{P_0, \cO\}$, which is disjoint from $\operatorname{Supp}(D)$ by construction. Thus, the codes $\CL(D, G)$ and $\CL(D, H)$ are well-defined with dimensions $k_G = \deg(G)$ and $k_H = \deg(H)$.

We first show that $\CL(D, H)$ is an MDS code. Suppose, for contradiction, that it is not. By Lemma \ref{lemma:MDS_characterization}, there exists a subset of $k_H$ distinct points in $\operatorname{Supp}(D)$ such that
\begin{equation}
\label{eq:sum(H)_1}
P_{i_1} \oplus \cdots \oplus P_{i_{k_H}} = \sigma(H) = [b]P_0.
\end{equation}
By grouping these $k_H$ points according to the cosets $\mathcal{H}_j$ to which they belong, we can rewrite the left-hand side of \eqref{eq:sum(H)_1} as
\begin{equation}
\label{eq:sum(H)_2}
P_{i_1} \oplus \cdots \oplus P_{i_{k_H}} = \bigoplus_{j=1}^s \bigoplus_{l=1}^{k_j} P_{j,l},
\end{equation}
where $P_{j,l} \in \mathcal{H}_j \cap \operatorname{Supp}(D)$ and $\sum_{j=1}^s k_j = k_H$ with $0 \le k_j \le m_j$.

Since each point $P_{j,l}$ belongs to $\mathcal{H}_j = [\lambda_j]P_0 \oplus \mathcal{H}$, the sum of $k_j$ such points evaluates to $[k_j \lambda_j]P_0 \oplus Q_j$ for some $Q_j \in \mathcal{H}$. Substituting this into \eqref{eq:sum(H)_1} and \eqref{eq:sum(H)_2} yields
\[
[b]P_0 = \bigoplus_{j=1}^s \big( [k_j \lambda_j]P_0 \oplus Q_j \big) = \left[ \sum_{j=1}^s k_j \lambda_j \right] P_0 \oplus Q',
\]
where $Q' = \bigoplus_{j=1}^s Q_j \in \mathcal{H}$. Rearranging terms, we obtain
\[
\left[ b - \sum_{j=1}^s k_j \lambda_j \right] P_0 = Q' \in \mathcal{H}.
\]
Since $\langle P_0 \rangle \cap \mathcal{H} = \{\cO\}$, this forces $b - \sum_{j=1}^s k_j \lambda_j \equiv 0 \pmod{\operatorname{ord}(P_0)}$,  contradicting Condition 2. Thus, $\CL(D, H)$ is MDS.

By an analogous argument, we show that $\CL(D, G)$ is MDS. Assume for contradiction that there exist $k_G = a+n-b$ points in $\operatorname{Supp}(D)$ summing to $\sigma(G) = [a]P_0$. Grouping these points by coset yields variables $\ell_j$ satisfying $0 \le \ell_j \le m_j$ and $\sum_{j=1}^s \ell_j = k_G$ such that
\[
[a]P_0 = \left[ \sum_{j=1}^s \ell_j \lambda_j \right] P_0 \oplus Q, \quad \mbox{for some } Q \in \mathcal{H}.
\]
This implies $\left[ a - \sum_{j=1}^s \ell_j \lambda_j \right] P_0 \in \mathcal{H}$, which contradicts Condition 3. Hence, $\CL(D, G)$ is also MDS.

Finally, we prove that $(\CL(D, G), \CL(D, H))$ is a linear complementary pair. From Corollary \ref{cor:LCP_elliptic}, we must verify that $\gcd(G,H)$ and $\lmd(G, H)-D$ are divisors of degree zero satisfying
\[
\sigma(\gcd(G,H)) \neq \cO \quad \mbox{and} \quad \sigma(\lmd(G,H)-D) \neq \cO.
\]
Since $0 < b-a < n$, it follows that $\gcd(G,H) = aP_0 - a\cO$, whose degree is $0$ and whose group sum is $\sigma(\gcd(G,H)) = [a]P_0$. As $a \not\equiv 0 \pmod{\operatorname{ord}(P_0)}$, we have $[a]P_0 \neq \cO$.

To evaluate $\sigma(\lmd(G,H)-D)$, we first determine $\sigma(D)$. The sum of all $m$ points in a single coset $\mathcal{H}_j$ is $[m \lambda_j]P_0 \oplus S_{\mathcal{H}}$, where $S_{\mathcal{H}} = \bigoplus_{Q \in \mathcal{H}} Q$. Since $P_0$ is removed from $\mathcal{H}_1$, we obtain
\[
\sigma(D) = \left( \bigoplus_{j=1}^s \left( [m \lambda_j]P_0 \oplus S_{\mathcal{H}} \right) \right) \ominus P_0 = [m S_\Lambda - 1]P_0 \oplus [s]S_{\mathcal{H}}.
\]
The divisor $\lmd(G,H) = bP_0 + (n-b)\cO$ has degree $n$, so $\lmd(G,H) - D$ has degree zero. Thus
\begin{align*}
\sigma(\lmd(G,H) - D) &= \sigma(\lmd(G,H)) \ominus \sigma(D) \\
&= [b]P_0 \ominus \big( [m S_\Lambda - 1]P_0 \oplus [s]S_{\mathcal{H}} \big) \\
&= [b - m S_\Lambda + 1]P_0 \ominus [s]S_{\mathcal{H}}.
\end{align*}
Consequently, $\sigma(\lmd(G, H)-D) = \cO$ if and only if
\[
[b - m S_\Lambda + 1]P_0 = [s]S_{\mathcal{H}} \in \cH.
\]
Since $\langle P_0 \rangle \cap \mathcal{H} = \{\cO\}$, Condition 1 guarantees that $[b - m S_\Lambda + 1]P_0 \neq \cO$. Thus, $\sigma(\lmd(G,H)-D) \neq \cO$, concluding the proof that $(\CL(D,G), \CL(D,H))$ is an LCP of MDS elliptic codes.
\end{proof}

\begin{corollary}
\label{cor:LCP_elliptic_single_coset}
Let $\mathcal{H} \le \cE(\F_q)$ be a subgroup of order $m$ and $P_0 \in \cE(\F_q)$ a rational point such that $\langle P_0 \rangle \cap \mathcal{H} = \{\cO\}$. Let $D = \sum_{P \in (P_0 \oplus \mathcal{H}) \setminus \{P_0\}} P$, so that $n = \deg(D) = m-1$.
Assume $a$ and $b$ are positive integers satisfying $0 < b-a < n$, such that
\[
    a, b \not\equiv 0 \pmod{\operatorname{ord}(P_0)} \quad \mbox{and} \quad b \not\equiv n \pmod{\operatorname{ord}(P_0)}.
\]
Then, the pair of codes
\[
    \big(\CL(D, aP_0+(n-b)\cO), \ \CL(D,bP_0-a\cO)\big)
\]
is an LCP of MDS elliptic codes with respective parameters $[n, a+n-b, b-a+1]_q$ and $[n, b-a, n-b+a+1]_q$.
\end{corollary}

\begin{proof}
We apply Theorem \ref{thm:LCP_elliptic_multicoset} for $s=1$. Here, $\Lambda = \{1\}$, so $\lambda_1 = 1$ and $S_\Lambda = 1$. 
Condition 1 becomes $b \not\equiv m - 1 \pmod{\operatorname{ord}(P_0)}$, that is, $b \not\equiv n \pmod{\operatorname{ord}(P_0)}$. 
For Condition 2, the only possible partition for $b-a$ is $k_1 = b-a$. The condition reduces to $b \not\equiv b-a \pmod{\operatorname{ord}(P_0)}$, which simplifies to $a \not\equiv 0 \pmod{\operatorname{ord}(P_0)}$. 
For Condition 3, the only partition for $a+n-b$ is $\ell_1 = a+n-b$. The condition reduces to $a \not\equiv a+n-b \pmod{\operatorname{ord}(P_0)}$, which simplifies to $b \not\equiv n \pmod{\operatorname{ord}(P_0)}$.
Since $a \not\equiv 0$ and $b \not\equiv n \pmod{\operatorname{ord}(P_0)}$ by hypothesis, all conditions of Theorem \ref{thm:LCP_elliptic_multicoset} are satisfied and the result follows.
\end{proof}

\section{Examples from Explicit Elliptic Curves}
\label{sec:explicit_examples}
In this section, we apply the general constructions established previously to explicitly specialize the LCPs of MDS elliptic codes over finite fields. By evaluating over specific elliptic curves with well-known cyclic or bi-cyclic group structures, such as Mordell curves and supersingular curves, we generate concrete families and numerical examples of these codes.

\begin{proposition}
\label{prop:mordell_p_5_or_11_mod_18}
Let $p$ be a prime satisfying $p \equiv 5$ or $11 \pmod{18}$ and set $n = (p-2)/3$. Let $\mathcal{E}: y^2 = x^3 + \lambda$ be a Mordell curve over $\mathbb{F}_p$ with $\lambda \in \mathbb{F}_p^*$. Let $P_0 \in \mathcal{E}(\mathbb{F}_p)$ be a rational point of order $3$, and let $\mathcal{H} \leq \mathcal{E}(\mathbb{F}_p)$ be the unique subgroup of order $n+1$. Define the evaluation divisor $D = \sum_{P \in (P_0 \oplus \mathcal{H}) \setminus \{P_0\}} P$. 
If $b$ is an integer such that $1 < b < n+1$, with $b \not\equiv 0 \pmod{3}$ and $b \not\equiv n \pmod 3$, then the pair of codes
\[
    \big(\CL(D, P_0+(n-b)\mathcal{O}), \ \CL(D,bP_0-\mathcal{O})\big)
\]
forms an LCP of MDS elliptic codes over $\F_p$ with respective parameters 
\[
    [n, n-b+1, b]_p \quad \text{and} \quad [n, b-1, n-b+2]_p.
\]
\end{proposition}

\begin{proof}
    First, observe that $p \equiv 5$ or $11 \pmod{18}$ implies $p \equiv 2 \pmod 3$. For a Mordell curve $y^2 = x^3 + \lambda$ over $\F_p$ with $p \equiv 2 \pmod 3$, the map $x \mapsto x^3$ is a bijection on $\F_p$. Consequently, the curve has exactly $|\cE(\F_p)| = p+1$ rational points. Moreover, the polynomial $x^3+\lambda$ has a unique root in $\F_p$, meaning the curve has exactly one rational point of order $2$. Since $|\cE(\F_p)| = p+1$ is even, this unique point of order $2$ ensures that the finite abelian group $\cE(\F_p)$ is cyclic.

    From $n = \frac{p-2}{3}$, we have $|\cE(\F_p)| = p+1 = 3(n+1)$. Since $\cE(\F_p)$ is cyclic, it contains a unique subgroup $\mathcal{H}$ of order $n+1$ and a unique subgroup of order $3$, which is generated by $P_0$. The condition $p \equiv 5$ or $11 \pmod{18}$ implies that $p \equiv 2$ or $5 \pmod 9$, which guarantees that $p+1$ is not divisible by $9$. Consequently, $3$ does not divide $n+1 = (p+1)/3$. Thus, $\gcd(|\mathcal{H}|, \operatorname{ord}(P_0)) = \gcd(n+1, 3) = 1$, which means that $\langle P_0 \rangle \cap \mathcal{H} = \{\cO\}$.

    By definition, the divisor $D$ is the sum of the points in the coset $P_0 \oplus \mathcal{H}$ excluding $P_0$. Therefore, $\deg(D) = |\mathcal{H}| - 1 = n$, and its support does not contain $P_0$. We now apply Corollary~\ref{cor:LCP_elliptic_single_coset} with $a=1$. The assumption $1 < b < n+1$ guarantees that $0 < b-1 < \deg(D)$. Furthermore, since $\operatorname{ord}(P_0) = 3$ and $\deg(D) = n$, by hypotheses, we obtain  $b \not\equiv 0 \pmod{\operatorname{ord}(P_0)}$ and $b \not\equiv \deg(D) \pmod{\operatorname{ord}(P_0)}$.  

    Since all conditions of Corollary~\ref{cor:LCP_elliptic_single_coset} are satisfied, the pair of codes forms an LCP of MDS elliptic codes over $\F_p$. Their respective dimensions are $k_G = n-b+1$ and $k_H = b-1$, completing the proof.
\end{proof}

\begin{proposition}
\label{prop:supersingular_x3_x_mod_8}
Let $p \equiv 3 \pmod 8$ be a prime and set $n = (p-3)/4$. Let $\mathcal{E}: y^2 = x^3 + x$ be an elliptic curve over $\mathbb{F}_p$. Let $P_0 \in \mathcal{E}(\mathbb{F}_p)$ be a rational point of order $4$ and let $\mathcal{H} \leq \mathcal{E}(\mathbb{F}_p)$ be the unique subgroup of order $n+1$. Define the evaluation divisor $D = \sum_{P \in (P_0 \oplus \mathcal{H}) \setminus \{P_0\}} P$.

If $b$ is an integer such that $1 < b < n+1$, with $b \not\equiv 0 \pmod{4}$ and $b \not\equiv n \pmod 4$, then the pair of codes
\[
    \big(\CL(D, P_0+(n-b)\mathcal{O}), \ \CL(D,bP_0-\mathcal{O})\big)
\]
forms an LCP of MDS elliptic codes over $\F_p$ with respective parameters 
\[
    [n, n-b+1, b]_p \quad \text{and} \quad [n, b-1, n-b+2]_p.
\]
\end{proposition}

\begin{proof}
    First, observe that the congruence $p \equiv 3 \pmod 8$ implies $p \equiv 3 \pmod 4$. It is a well-known fact that the elliptic curve $y^2 = x^3 + x$ over $\F_p$ with $p \equiv 3 \pmod 4$ has exactly $|\cE(\F_p)| = p+1$ rational points. Because $p \equiv 3 \pmod 4$, $-1$ is a quadratic non-residue modulo $p$. Therefore, the polynomial $x^3+x = x(x^2+1)$ has a unique root in $\F_p$, meaning the curve has exactly one rational point of order $2$. Since $|\cE(\F_p)| = p+1$ is even, this unique point of order $2$ ensures that the finite abelian group $\cE(\F_p)$ is cyclic.

    From $n = (p-3)/4$, we have $|\cE(\F_p)| = p+1 = 4(n+1)$. Since $\cE(\F_p)$ is cyclic, it contains a unique subgroup $\mathcal{H}$ of order $n+1$ and a unique subgroup of order $4$, which is generated by $P_0$. Furthermore, the condition $p \equiv 3 \pmod 8$ implies that $p+1 \equiv 4 \pmod 8$, meaning that $n+1 = (p+1)/4$ is an odd integer. Thus, $\gcd(|\mathcal{H}|, \operatorname{ord}(P_0)) = \gcd(n+1, 4) = 1$, which means that $\langle P_0 \rangle \cap \mathcal{H} = \{\cO\}$.

    By definition, the divisor $D$ is the sum of the points in the coset $P_0 \oplus \mathcal{H}$ excluding $P_0$. Therefore, $\deg(D) = |\mathcal{H}| - 1 = n$, and its support does not contain $P_0$. We now apply Corollary~\ref{cor:LCP_elliptic_single_coset} with $a=1$. The assumption $1 < b < n+1$ guarantees that $0 < b-1 < \deg(D)$. Furthermore, since $\operatorname{ord}(P_0) = 4$ and $\deg(D) = n$, the hypotheses $b \not\equiv 0 \pmod 4$ and $b \not\equiv n \pmod 4$ translate directly to $b \not\equiv 0 \pmod{\operatorname{ord}(P_0)}$ and $b \not\equiv \deg(D) \pmod{\operatorname{ord}(P_0)}$. 

    Since all conditions of Corollary~\ref{cor:LCP_elliptic_single_coset} are satisfied, the pair of codes forms an LCP of MDS elliptic codes over $\F_p$. Their respective dimensions are $k_G = n-b+1$ and $k_H = b-1$, completing the proof.
\end{proof}

\begin{example}
\label{ex:mordell_p_5_or_11_mod_18}
Let $p=47 \equiv 11 \pmod{18}$, which yields $n = \frac{47-2}{3} = 15$. For a Mordell curve $\mathcal{E}: y^2 = x^3 + \lambda$ (with $\lambda \neq 0$) over $\mathbb{F}_{47}$, Proposition~\ref{prop:mordell_p_5_or_11_mod_18} provides explicit LCPs of MDS codes of length $15$ using a subgroup $\mathcal{H}$ of order $16$ and a rational point $P_0$ of order $3$.

The integer $b$ must satisfy $1 < b < 16$, $b \not\equiv 0 \pmod 3$ and $b \not\equiv 15 \equiv 0 \pmod 3$. Thus, $b \in \{2, 4, 5, 7, 8, 10, 11, 13, 14\}$. For each choice of $b$, the construction yields a pair of MDS codes $\CL(D,G)$ and $\CL(D,H)$ with dimensions $16-b$ and $b-1$, respectively. Their parameters are summarized below:
\[
\begin{array}{c|c|c}
b & \CL(D,G) & \CL(D,H) \\
\hline
2  & [15,14, 2]_{47} & [15, 1,15]_{47} \\
4  & [15,12, 4]_{47} & [15, 3,13]_{47} \\
5  & [15,11, 5]_{47} & [15, 4,12]_{47} \\
7  & [15, 9, 7]_{47} & [15, 6,10]_{47} \\
8  & [15, 8, 8]_{47} & [15, 7, 9]_{47} \\
10 & [15, 6,10]_{47} & [15, 9, 7]_{47} \\
11 & [15, 5,11]_{47} & [15,10, 6]_{47} \\
13 & [15, 3,13]_{47} & [15,12, 4]_{47} \\
14 & [15, 2,14]_{47} & [15,13, 3]_{47} 
\end{array}
\]
\end{example}

\begin{example}
\label{ex:supersingular_x3_x_lcp}
Let $p=59 \equiv 3 \pmod 8$, which gives $n = \frac{59-3}{4} = 14$. For the curve $\mathcal E:y^2=x^3+x$ over $\mathbb F_{59}$, Proposition~\ref{prop:supersingular_x3_x_mod_8} provides explicit LCPs of length $14$ using a subgroup $\mathcal H$ of order $15$ and a rational point $P_0$ of order $4$.

The integer $b$ must satisfy $1 < b < 15$ and the congruences $b \not\equiv 0\pmod 4 $ and $b \not\equiv 14 \equiv 2 \pmod 4$. This implies that $b \in \{3,5,7,9,11,13\}$. For each choice of $b$, the construction yields an LCP of MDS codes $\CL(D,G)$ and $\CL(D,H)$ with dimensions $15-b$ and $b-1$, respectively. Their parameters are summarized below:
\[
\begin{array}{c|c|c}
b & \CL(D,G) & \CL(D,H) \\
\hline
3 & [14,12,3]_{59} & [14,2,13]_{59} \\
5 & [14,10,5]_{59} & [14,4,11]_{59} \\
7 & [14,8,7]_{59}  & [14,6,9]_{59}  \\
9 & [14,6,9]_{59}  & [14,8,7]_{59}  \\
11 & [14,4,11]_{59} & [14,10,5]_{59} \\
13 & [14,2,13]_{59} & [14,12,3]_{59} 
\end{array}
\]
\end{example}

While the preceding results focus on the single-coset case ($s=1$), the following example illustrates the general multi-coset construction ($s>1$) established in Theorem~\ref{thm:LCP_elliptic_multicoset}.

\begin{example}
\label{ex:multicoset_s2_45_percent}
Consider the supersingular elliptic curve $\mathcal{E}: y^2 = x^3 + x$ over $\mathbb{F}_{19}$. Since $19 \equiv 3 \pmod 4$, we have $|\mathcal{E}(\mathbb{F}_{19})| = 20$ and  $\mathcal{E}(\mathbb{F}_{19}) \cong \mathbb{Z}/20\mathbb{Z}$ is cyclic. 
Let $P_0 \in \mathcal{E}(\mathbb{F}_{19})$ be a rational point of order $4$ and $\mathcal{H}$ be the unique subgroup of order $m=5$;  in particular, $\langle P_0 \rangle \cap \mathcal{H} = \{\mathcal{O}\}$. We consider $\Lambda = \{1, 3\}$ to form $s=2$ cosets. The resulting evaluation divisor $D$ has degree $n = 2 \cdot 5 - 1 = 9$.

Setting $a = 1$, the integer $b$ is restricted to $1 < b < 10$ and $b \not\equiv 0 \pmod 4$. We verify the conditions of Theorem~\ref{thm:LCP_elliptic_multicoset}:
\begin{enumerate}[label={\tiny\textbullet}]
    \item Condition 1 requires $b \not\equiv m S_\Lambda - 1 \pmod 4$. Since $S_\Lambda = 4$, this gives $b \not\equiv 5(4) - 1 \equiv 3 \pmod 4$, restricting the allowed values to $b \in \{2, 5, 6, 9\}$.
    
    \item Condition 2 states that for any sequence of non-negative integers $(k_1, k_2)$ such that $k_1+k_2 = b-1$, we must have $b \not\equiv k_1 + 3k_2 \pmod 4$. Substituting $k_2 = b-1-k_1$, the congruence reduces to $2b \not\equiv 1-2k_1 \pmod 4$. Since $2b$ is even and $1-2k_1$ is odd, this holds trivially for all $b$.
    
    \item Condition 3 determines $1 \not\equiv \ell_1 + 3\ell_2 \pmod 4$ for any sequence of non-negative integers $(\ell_1, \ell_2)$ such that $\ell_1+\ell_2=10-b$. This similarly simplifies to $b \not\equiv 3+2\ell_1 \pmod 4$. This holds for $b \in \{2, 6\}$ (which are even) but fails for $b \in \{5, 9\}$ (which are odd) upon choosing $\ell_1=1$.
\end{enumerate}
Consequently, $b \in \{2, 6\}$. By Theorem~\ref{thm:LCP_elliptic_multicoset}, this yields the following LCPs of MDS elliptic codes of length $9$:
\[
\begin{array}{c|c|c}
b & \CL(D,G) & \CL(D,H) \\
\hline
2 & [9,8,2]_{19} & [9,1,9]_{19} \\
6 & [9,4,6]_{19} & [9,5,5]_{19} \\
\end{array}
\]
\end{example}

Finally, we apply the multi-coset approach to construct large-length LCPs of MDS codes over quadratic extensions. Using the $\Z_{p+1} \oplus \Z_{p+1}$ group structure of supersingular curves over $\F_{p^2}$, we generate the following infinite family by taking the union of $s$ cosets.

\begin{proposition}
\label{cor:large_length_supersingular_LCP}
Let $p \ge 7$ be a prime satisfying $p \equiv 3 \pmod 4$. Let $\mathcal{E}: y^2 = x^3 - x$ be an elliptic curve over $\mathbb{F}_{p^2}$. Let $\mathcal{H}$ and $\langle P_0 \rangle$ be disjoint subgroups of $\mathcal{E}(\F_{p^2})$ of order $p+1$. 
For an integer $1 \le s \le (p-1)/2$, define   
\[
U=\left(\bigcup_{j=1}^s [j]P_0 \oplus \cH\right) \setminus \{ P_0 \} \quad \mbox{and} \quad D = \sum_{P \in U} P.
\]
Let $n = \deg(D) = s(p+1) - 1$. If $k \ge 1$ is an integer satisfying $k(2s - 1) + 2 \le p$, define $a = k(s-1)+1$ and $b = ks+1$. Then, the pair of codes
\[
    \big(\CL(D, aP_0+(n-b)\mathcal{O}), \ \CL(D,bP_0-a\mathcal{O})\big)
\]
forms an LCP of MDS elliptic codes over $\F_{p^2}$ with respective parameters
\[
    [n, n-k, k+1]_{p^2} \quad \text{and} \quad [n, k, n-k+1]_{p^2}.
\]
\end{proposition}

\begin{proof}
    First, recall that since $p \equiv 3 \pmod 4$, the curve $\cE$ over $\F_{p}$ is supersingular with trace of Frobenius zero. Standard elliptic curve theory over finite fields (see, e.g., \cite[Chapter 4]{Washington2008}) dictates that $\cE(\F_{p^2}) \cong \Z_{p+1} \oplus \Z_{p+1}$. This guarantees the existence of the disjoint subgroups $\mathcal{H}$ and $\langle P_0 \rangle$, both of order $p+1$. We apply Theorem \ref{thm:LCP_elliptic_multicoset} with 
    \[
    \operatorname{ord}(P_0) = p+1, \quad \Lambda = \{1, \dots, s\}, \quad m_1 = p \ \mbox{ and } \ m_j = p+1 \mbox{ for } 2 \le j \le s.
    \]
    The condition $k \ge 1$ implies $a = k(s-1) + 1 \ge 1$, which ensures $0 < a < b < p+1$. Consequently, $0 < b-a = k < n$ and $a, b \not\equiv 0 \pmod{p+1}$.

    We now verify the three conditions of Theorem \ref{thm:LCP_elliptic_multicoset}.
    
For Condition 1, $m S_\Lambda - 1 = (p+1)S_\Lambda - 1 \equiv -1 \equiv p \pmod{p+1}$. Since $b < p$, we have $b \not\equiv m S_\Lambda - 1 \pmod{p+1}$.

For Condition 2, consider a sequence $k_j$ such that $\sum_{j=1}^s k_j = b-a = k$.  The sum $\Sigma_k = \sum_{j=1}^s j k_j$ satisfies $k \le \Sigma_k \le ks$. Since $b = ks + 1$, we have $b > \Sigma_k$, yielding $b \not\equiv \Sigma_k \pmod{p+1}$.

For Condition 3, consider a sequence $\ell_j$ such  that $\sum_{j=1}^s \ell_j=a+n-b = n - k$. Setting $v_1 = p - \ell_1$ and $v_j = p+1 - \ell_j$ for $j \ge 2$, we obtain $\sum_{j=1}^s v_j = k$. The sum $\Sigma_\ell = \sum_{j=1}^s j \ell_j$ evaluates to

\[
    \Sigma_\ell = p + (p+1)\sum_{j=2}^s j - \sum_{j=1}^s j v_j \equiv -1 - \Sigma_v \pmod{p+1},
\]
where $\Sigma_v = \sum_{j=1}^s j v_j$. Since $\sum_{j=1}^s v_j = k$, we have $k \le \Sigma_v \le ks$. Thus, $\Sigma_\ell \pmod{p+1}$ takes values in the interval $[p-ks, p-k]$. Finally, since $a = ks - k + 1$ and $2ks - k + 2 \le p$, we have $a \le p - ks - 1 < p - ks$. Hence, $a \not\equiv \Sigma_\ell \pmod{p+1}$.

Since all conditions of Theorem \ref{thm:LCP_elliptic_multicoset} are satisfied, the pair of codes forms the desired LCP of MDS elliptic codes, completing the proof.
\end{proof}

To illustrate the possible values of the parameters and the geometric conditions of this family, we apply the construction explicitly to $p=19$, considering all possible choices of cosets and dimensions.

\begin{example}
\label{ex:large_length_p19_full}
Let $p = 19 \equiv 3 \pmod 4$. We consider the supersingular elliptic curve $\cE: y^2 = x^3 - x$ over $\F_{361}$. The group of rational points is isomorphic to $\Z_{20} \oplus \Z_{20}$.  Let $\cH$ and $\langle P_0 \rangle$ be disjoint subgroups of $\cE(\F_{361})$ of order $p+1 = 20$. The number of cosets $s$ can take values from $1$ to $(p-1)/2 = 9$. For a given $s$, the evaluation divisor $D$ is supported on $n = 20s - 1$ rational points.

By Proposition \ref{cor:large_length_supersingular_LCP}, for each $s \in \{1, \dots, 9\}$, the parameter $k$ must be a positive integer satisfying $k(2s - 1) + 2 \le 19$, so $k \le \lfloor 17 / (2s - 1) \rfloor$. Each valid pair $(s, k)$ determines the coefficients $a = k(s-1)+1$ and $b = ks+1$. 

By setting $G = aP_0 + (n-b)\cO$ and $H = bP_0 - a\cO$, we obtain an LCP of MDS elliptic codes over $\F_{361}$ with dimensions $n-k$ and $k$. The parameters of the resulting codes are summarized below: 
\[
\begin{array}{c|c|c|c|c}
s & n & k & \CL(D,G) & \CL(D,H) \\
\hline
1 & 19 & 1 \le k \le 17 & [19, 19-k, k+1]_{361} & [19, k, 20-k]_{361} \\
2 & 39 & 1 \le k \le 5  & [39, 39-k, k+1]_{361} & [39, k, 40-k]_{361} \\
3 & 59 & 1 \le k \le 3  & [59, 59-k, k+1]_{361} & [59, k, 60-k]_{361} \\
4 & 79 & 1 \le k \le 2  & [79, 79-k, k+1]_{361} & [79, k, 80-k]_{361} \\
5 & 99 & 1 & [99, 98, 2]_{361} & [99, 1, 99]_{361} \\
6 & 119 & 1 & [119, 118, 2]_{361} & [119, 1, 119]_{361} \\
7 & 139 & 1 & [139, 138, 2]_{361} & [139, 1, 139]_{361} \\
8 & 159 & 1 & [159, 158, 2]_{361} & [159, 1, 159]_{361} \\
9 & 179 & 1 & [179, 178, 2]_{361} & [179, 1, 179]_{361}
\end{array}
\]
As shown above, increasing the number of cosets $s$ results in longer lengths $n$, but restricts the possible values of $k$. In particular, for $s \ge 5$, the bounding condition forces $k=1$, yielding component codes of dimensions $1$ and $n-1$.
\end{example}

\section{Conclusion}
\label{sec:conclusion}

In this paper, we investigated the theoretical limitations and explicit constructions of LCPs of MDS elliptic codes over finite fields. By applying the Abel-Jacobi theorem, we established sufficient geometric conditions for LCPs of elliptic codes and proved a fundamental non-existence result: an LCP of elliptic codes cannot consist entirely of MDS codes if evaluated over all rational points. To overcome this geometric obstruction, we developed a general multi-coset algebraic construction that systematically breaks the zero-sum relation. Relying on the precise group structure of the underlying curve, this approach enabled the derivation of exact parameters for explicit infinite families of LCPs of MDS codes from Mordell and supersingular curves. All explicit families and computational parameters have been verified using MAGMA. For future work, it would be interesting to extend this geometric framework to construct LCPs of optimal codes by exploiting the group of divisors of algebraic curves of higher genus.

\printbibliography 

\end{document}